\documentclass[a4paper]{aa}
\usepackage[T1]{fontenc}
\usepackage[utf8]{inputenc}
\usepackage{graphicx}
\usepackage{color}
\usepackage{txfonts}
\usepackage{microtype}
\usepackage{ellipsis}
\usepackage{natbib}
\bibpunct{(}{)}{;}{a}{}{,}
\usepackage[pdftex,colorlinks=true,urlcolor=blue]{hyperref}
\usepackage[xindy,nonumberlist,nopostdot,nogroupskip]{glossaries}

\makeglossaries
\usepackage[xindy]{imakeidx}
\makeindex

\begin{document} 
\title{Navigation in the Ancient Mediterranean and Beyond}
\author{M. Nielbock}
\institute{Haus der Astronomie, Campus MPIA, Königstuhl 17, D-69117 Heidelberg, Germany\\
              \email{nielbock@hda-hd.de}}
\date{Received August 8, 2016; accepted March 2, 2017}

\abstract{This lesson unit has been developed within the framework the EU Space Awareness project. It provides an insight into the history and navigational methods of the Bronze Age Mediterranean peoples. The students explore the link between exciting history and astronomical knowledge. Besides an overview of ancient seafaring in the Mediterranean, the students explore in two hands-on activities early navigational skills using the stars and constellations and their apparent nightly movement across the sky. In the course of the activities, they become familiar with the stellar constellations and how they are distributed across the northern and southern sky.}


\maketitle
%

\section{Background information}

\subsection{Cardinal directions}
The cardinal directions are defined by astronomical processes like diurnal and 
annual apparent movements of the Sun and the apparent movements of the stars. In 
ancient and prehistoric times, the sky certainly had a different significance 
than today. This is reflected in the many myths all around the world. As a 
result, we can assume that the processes in the sky have been watched and 
monitored closely. In doing this, the underlying cycles and visible phenomena 
were easy to observe.

\begin{figure}[!ht]
 \centering
 \resizebox{0.75\hsize}{!}{\includegraphics{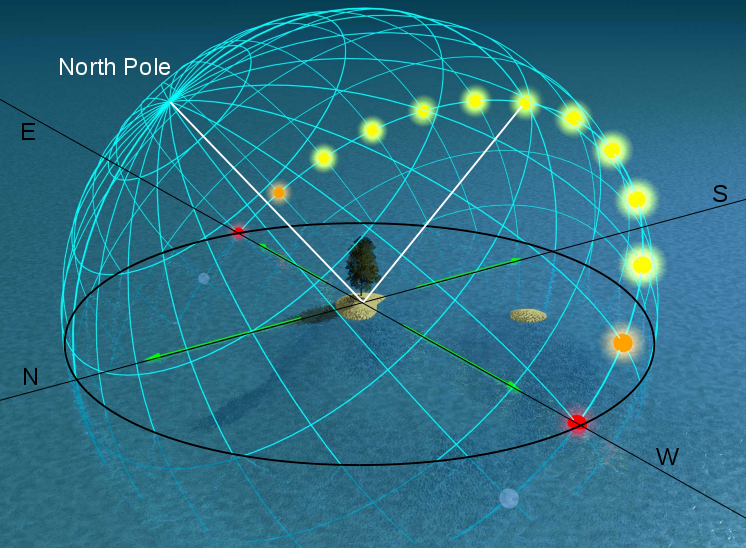}}
 \caption{Apparent diurnal movement of the Sun in the northern hemisphere at 
equinox. The Sun reaches its highest elevation above the horizon to the south. 
In the southern hemisphere, the Sun culminates to the north (Tau'olunga, 
\url{https://commons.wikimedia.org/wiki/File:Equinox-50.jpg}, ``Equinox-50'', 
horizontal coordinate system and annotations added by Markus Nielbock, 
\url{https://creativecommons.org/licenses/by-sa/3.0/legalcode}).}
 \label{f:cardinal}
\end{figure}

For any given position on Earth except the equatorial region, the Sun always 
culminates towards the same direction (Fig.~\ref{f:cardinal}). The region 
between the two tropics 23\fdg5 north and south of the equator is special, 
because the Sun can attain zenith positions at local noon throughout the year. 
During night, the stars rotate around the celestial poles. Archaeological 
evidence from prehistoric eras like burials and the orientation of buildings 
demonstrates that the cardinal directions were common knowledge in a multitude 
of cultures already many millennia ago. Therefore, it is obvious that they were 
applied to early navigation. The magnetic compass was unknown in Europe until 
the 13th century CE \citep{lane_economic_1963}.

\subsection{Latitude and longitude}
\begin{figure}[!ht]
 \resizebox{\hsize}{!}{\includegraphics{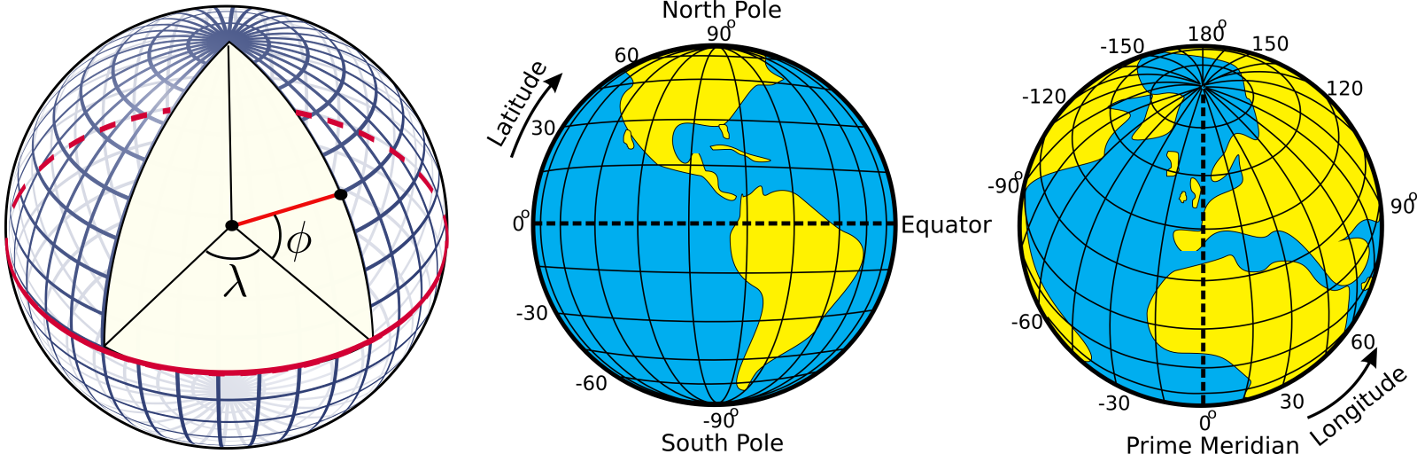}}
 \caption{Illustration of how the latitudes and longitudes of the Earth are 
defined (Peter Mercator, djexplo, CC0).}
 \label{f:latlong}
\end{figure}

Any location on an area is defined by two coordinates. The surface of a sphere 
is a curved area, but using coordinates like up and down does not make much 
sense, because the surface of a sphere has neither a beginning nor an ending. 
Instead, we can use spherical polar coordinates originating from the centre of 
the sphere with the radius being fixed (Fig.~\ref{f:latlong}). Two angular 
coordinates remain. Applied to the Earth, they are called the latitude and the 
longitude. Its rotation provides the symmetry axis. The North Pole is defined as 
the point, where the theoretical axis of rotation meets the surface of the 
sphere and the rotation is counter-clockwise when looking at the North Pole from 
above. The opposite point is the South Pole. The equator is defined as the great 
circle half way between the two poles.

The latitudes are circles parallel to the equator. They are counted from 
$0\degr$ at the equator to $\pm 90\degr$ at the poles. The longitudes are great 
circles connecting the two poles of the Earth. For a given position on Earth, 
the longitude going through the zenith, the point directly above, is called the 
meridian. This is the line the Sun apparently crosses at local noon. The origin 
of this coordinate is defined as the Prime Meridian, and passes Greenwich, where 
the Royal Observatory of England is located. From there, longitudes are counted 
from $0\degr$ to $\pm 180\degr$.

Example: Heidelberg in Germany is located at 49\fdg4 North and 8\fdg7 East.

\subsection{Elevation if the pole (pole height)}
\begin{figure}
 \resizebox{\hsize}{!}{\includegraphics{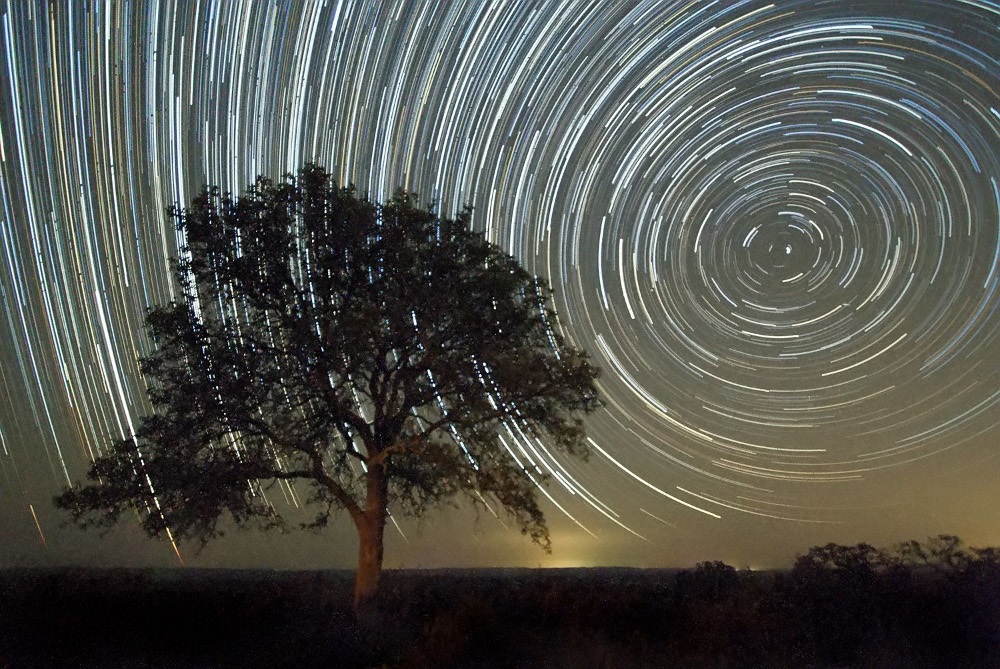}}
 \caption{Trails of stars at the sky after an exposure time of approximately 2 
hours (Ralph Arvesen, Live Oak star trails, 
\url{https://www.flickr.com/photos/rarvesen/9494908143}, 
\url{https://creativecommons.org/licenses/by/2.0/legalcode}).}
 \label{f:trails}
\end{figure}

If we project the terrestrial coordinate system of latitudes and longitudes at 
the sky, we get the celestial equatorial coordinate system. The Earth's equator 
becomes the celestial equator and the geographic poles are extrapolated to build 
the celestial poles. If we were to make a photograph with a long exposure of the 
northern sky, we would see from the trails of the stars that they all revolve 
about a common point, the northern celestial pole (Fig.~\ref{f:trails}).

\begin{figure}[!ht]
 \centering
 \resizebox{0.5\hsize}{!}{\includegraphics{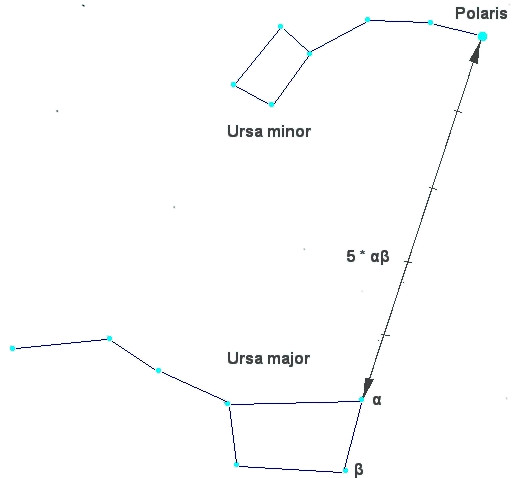}}
 \caption{Configuration of the two constellations Ursa Major (Great Bear) and 
Ursa Minor (Little Bear) in the northern sky. Polaris, the North Star, which is 
close to the true celestial north pole, is the brightest star in Ursa Minor 
(Bonč, 
\url{
https://commons.wikimedia.org/wiki/File:Ursa_Major_-_Ursa_Minor_-_Polaris.jpg}, 
``Ursa Major – Ursa Minor – Polaris'', colours inverted by Markus Nielbock, 
\url{https://creativecommons.org/licenses/by-sa/3.0/legalcode}).}
 \label{f:polaris}
\end{figure}

In the northern hemisphere, there is a moderately bright star near the celestial 
pole, the North Star or Polaris. It is the brightest star in the constellation 
the Little Bear, Ursa Minor (Fig.~\ref{f:polaris}). In our era, Polaris is less 
than a degree off. However, 1000 years ago, it was $8\degr$ away from the pole. 
Therefore, today we can use it as a proxy for the position of the celestial 
North Pole. At the southern celestial pole, there is no such star that can be 
observed with the naked eye. Other procedures have to be applied to find it.

\begin{figure}
 \centering
 \resizebox{0.45\hsize}{!}{\includegraphics{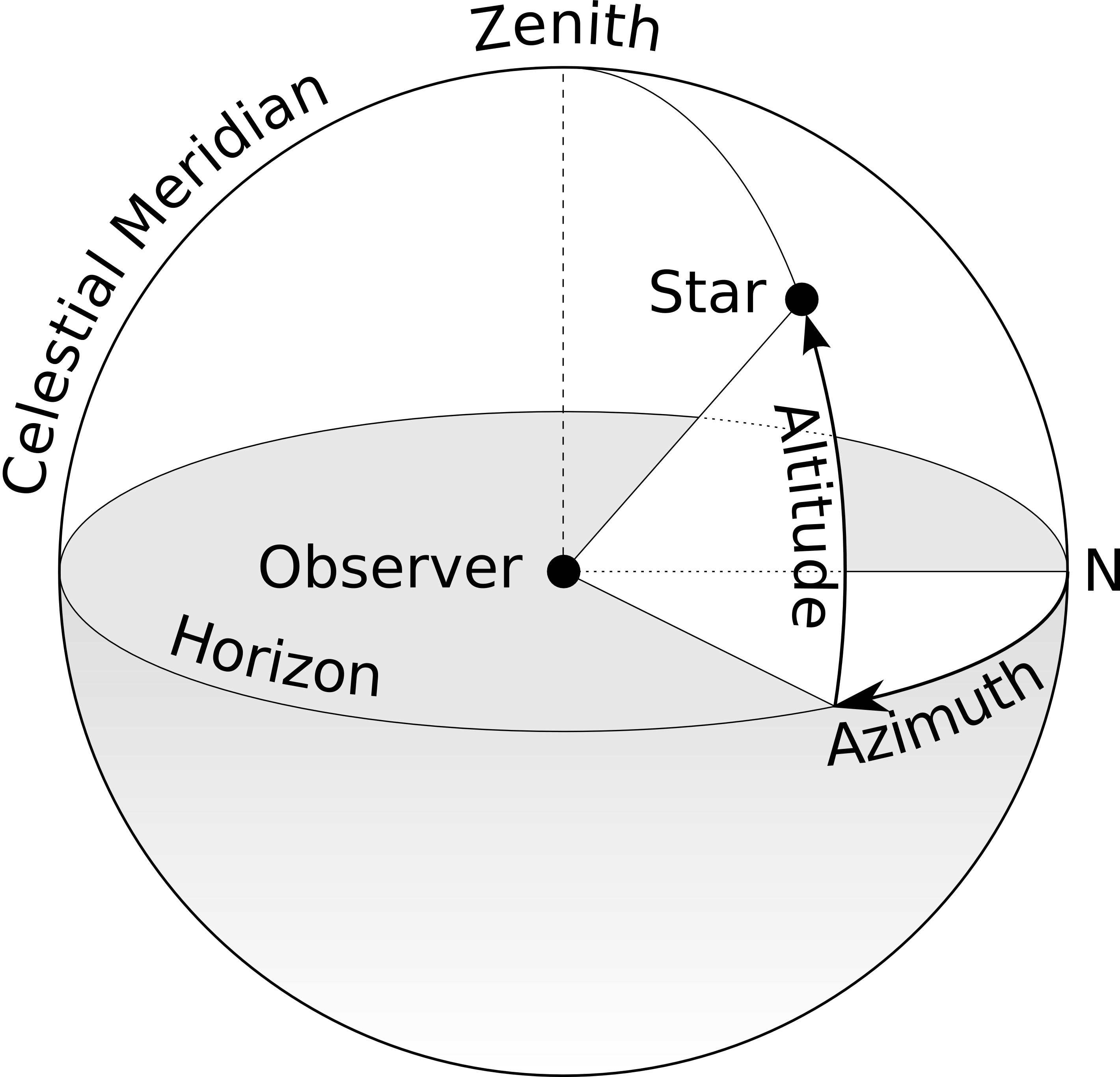}}
 \caption{Illustration of the horizontal coordinate system. The observer is the 
origin of the coordinates assigned as azimuth and altitude or elevation 
(TWCarlson, \url{
https://commons.wikimedia.org/wiki/File:Azimuth-Altitude_schematic.svg},
``Azimuth-Altitude 
schematic'', \url{https://creativecommons.org/licenses/by-sa/3.0/legalcode}).}
  \label{f:altaz}
\end{figure}

If we stood exactly at the geographic North Pole, Polaris would always be 
directly overhead. We can say that its elevation would be (almost) $90\degr$. 
This information already introduces the horizontal coordinate system 
(Fig.~\ref{f:altaz}). It is the natural reference we use every day. We, the 
observers, are the origin of that coordinate system located on a flat plane 
whose edge is the horizon. The sky is imagined as a hemisphere above. The angle 
between an object in the sky and the horizon is the altitude or elevation. The 
direction within the plane is given as an angle between $0\degr$ and $360\degr$, 
the azimuth, which is usually counted clockwise from north. In navigation, this 
is also called the bearing. The meridian is the line that connects North and 
South at the horizon and passes the zenith.

\begin{figure}[!ht]
\centering
 \resizebox{0.9\hsize}{!}{\includegraphics{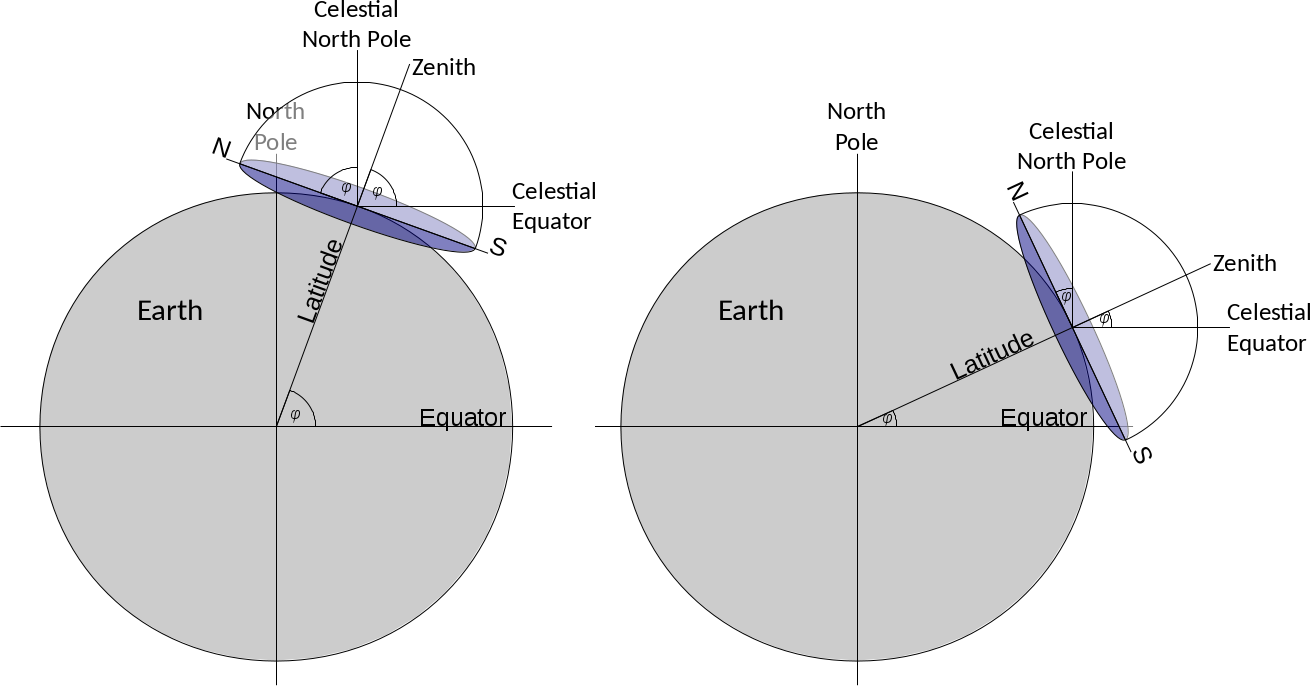}}
 \caption{When combining the three coordinate systems (terrestrial spherical, 
celestial equatorial, local horizontal), it becomes clear that the latitude of 
the observer is exactly the elevation of the celestial pole, also known as the 
pole height (own work).}
   \label{f:poleheight}
\end{figure}

\begin{figure*}
\centering
 \resizebox{\hsize}{!}{\includegraphics{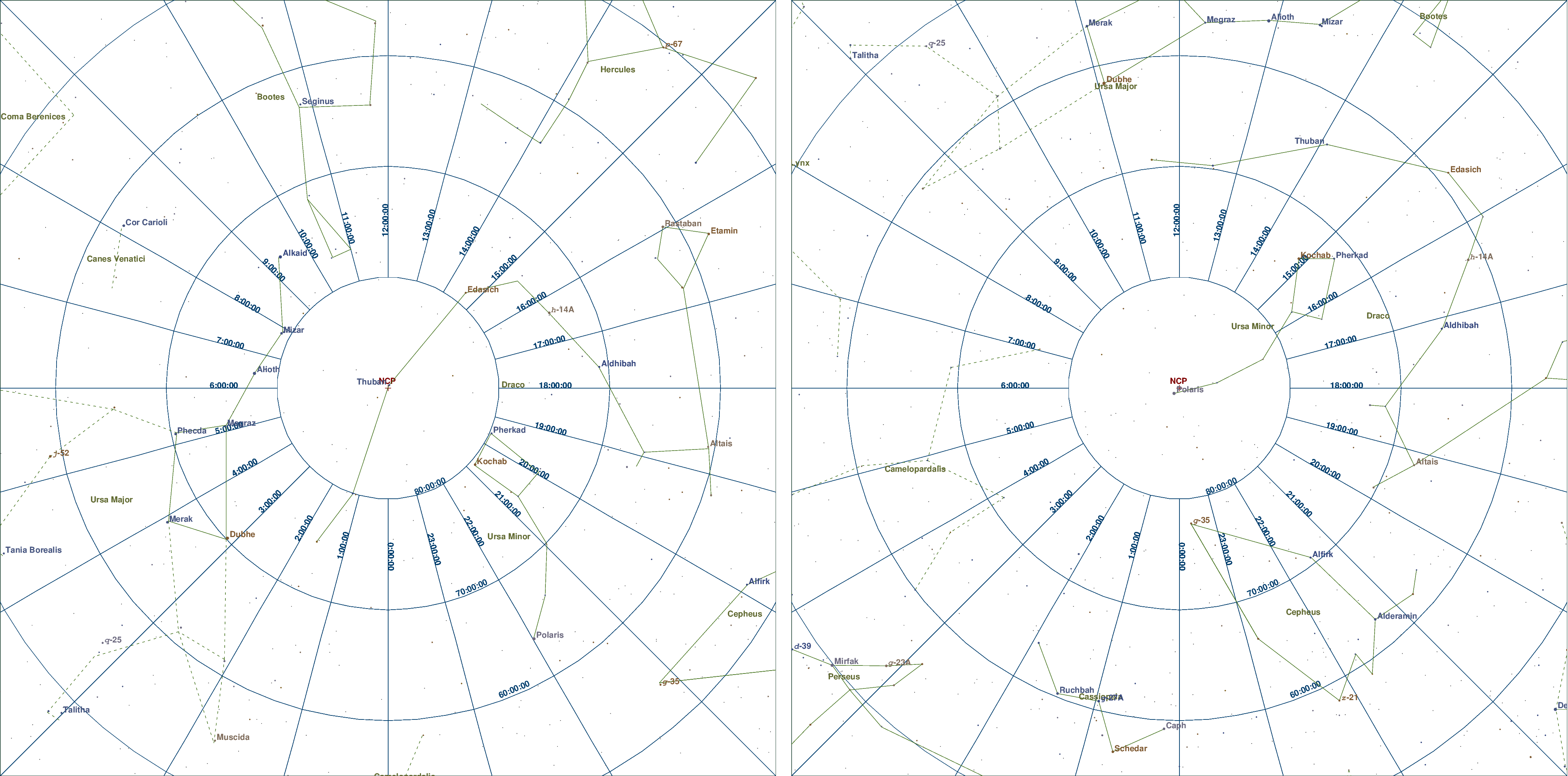}}
 \caption{Star charts of the northern celestial pole region for the years 2750 
BCE and 2016 CE (own work, created with XEphem Version 3.7.6 produced by Elwood 
C. Downey and distributed by the Clear Sky Institute Inc., Solon, Iowa, USA, 
\url{http://www.xephem.com}).}
 \label{f:hemisphere}
\end{figure*}

For any other position on Earth, the celestial pole or Polaris would appear at 
an elevation smaller than $90\degr$. At the equator, it would just graze the 
horizon, i.e. be at an elevation of $0\degr$. The correlation between the 
latitude (North Pole = $90\degr$, Equator = $0\degr$) and the elevation of 
Polaris is no coincidence. Figure~\ref{f:poleheight} combines all three 
mentioned coordinate systems. For a given observer at any latitude on Earth, the 
local horizontal coordinate system touches the terrestrial spherical polar 
coordinate system at a single tangent point. The sketch demonstrates that the 
elevation of the celestial North Pole, also called the pole height, is exactly 
the northern latitude of the observer on Earth. From this we can conclude that 
if we measure the elevation of Polaris, we can determine our latitude on Earth 
with reasonable precision.

\subsection{Circumpolar stars and constellations}
In ancient history, e.g. during the Bronze Age, Polaris could not be used to 
determine north. Due to the precession of the Earth's axis, it was about 
$30\degr$ away from the celestial North Pole in 3,500 BCE. Instead, the star 
Thuban ($\alpha$~Draconis) was more appropriate, as it was less than $4\degr$ 
off. However, it was considerably fainter than Polaris and perhaps not always 
visible to the naked eye.

When looking at the night sky, some stars within a certain radius around the 
celestial poles never set; they are circumpolar (see Fig.~\ref{f:trails}). 
Navigators were skilled enough to determine the true position of the celestial 
pole, by observing a few stars close to it. This method also works for the 
southern celestial pole. There are two videos that demonstrate the phenomenon.

\medskip\noindent
CircumpolarStars Heidelberg 49degN (Duration: 0:57)\\
\url{https://youtu.be/uzeey9VPA48}

\medskip\noindent
CircumpolarStars Habana 23degN (Duration: 0:49)\\
\url{https://youtu.be/zggfQC_d7UQ}

\medskip
They show the movement of the night sky when looking north for two different 
latitudes coinciding with the cities of Heidelberg, Germany ($49\degr$ North) 
and Lisbon, Portugal ($23\degr$ North). The videos illustrate that

\begin{enumerate}
\item there are always stars and constellations that never set. Those are the 
circumpolar stars and constellations.
\item the angle between the celestial pole (Polaris) and the horizon depends on 
the latitude of the observer. In fact, these angles are identical.
\item the circumpolar region depends on the latitude of the observer. It is 
bigger for locations closer to the pole.
\end{enumerate}

If the students are familiar with the usage of a planisphere, they can study the 
same phenomenon by watching the following two videos. They show the rotation of 
the sky for the latitudes $20\degr$ and $4\degr$. The transparent area reveals 
the visible sky for a given point in time. The dashed circle indicates the 
region of circumpolar stars and constellations.

\medskip\noindent
CircumPolarStars phi N20 (Duration: 0:37)\\
\url{https://youtu.be/Uv-xcdqhV00}

\medskip\noindent
CircumPolarStars phi N45 (Duration: 0:37)\\
\url{https://youtu.be/VZ6RmdzbpPw}

\medskip
When sailing north or south, sailors observe that with changing elevation of the 
celestial pole the circumpolar range is altered, too. Therefore, whenever 
navigators see the same star or constellation culminating – i.e. passing the 
meridian – at the same elevation, they stay on the ``latitude''.  Although the 
educated among the ancient Greek were familiar with the concept of latitude of a 
spherical Earth, common sailors were probably not. For them, it was sufficient 
to realise the connection between the elevation of stars and their course. 
Ancient navigators knew the night sky very well. In particular, they utilised 
the relative positions of constellation which helped them to determine their 
position in terms of latitude.

\subsection{Early seafaring and navigation in the Mediterranean}
\begin{figure}
\centering
 \resizebox{\hsize}{!}{\includegraphics{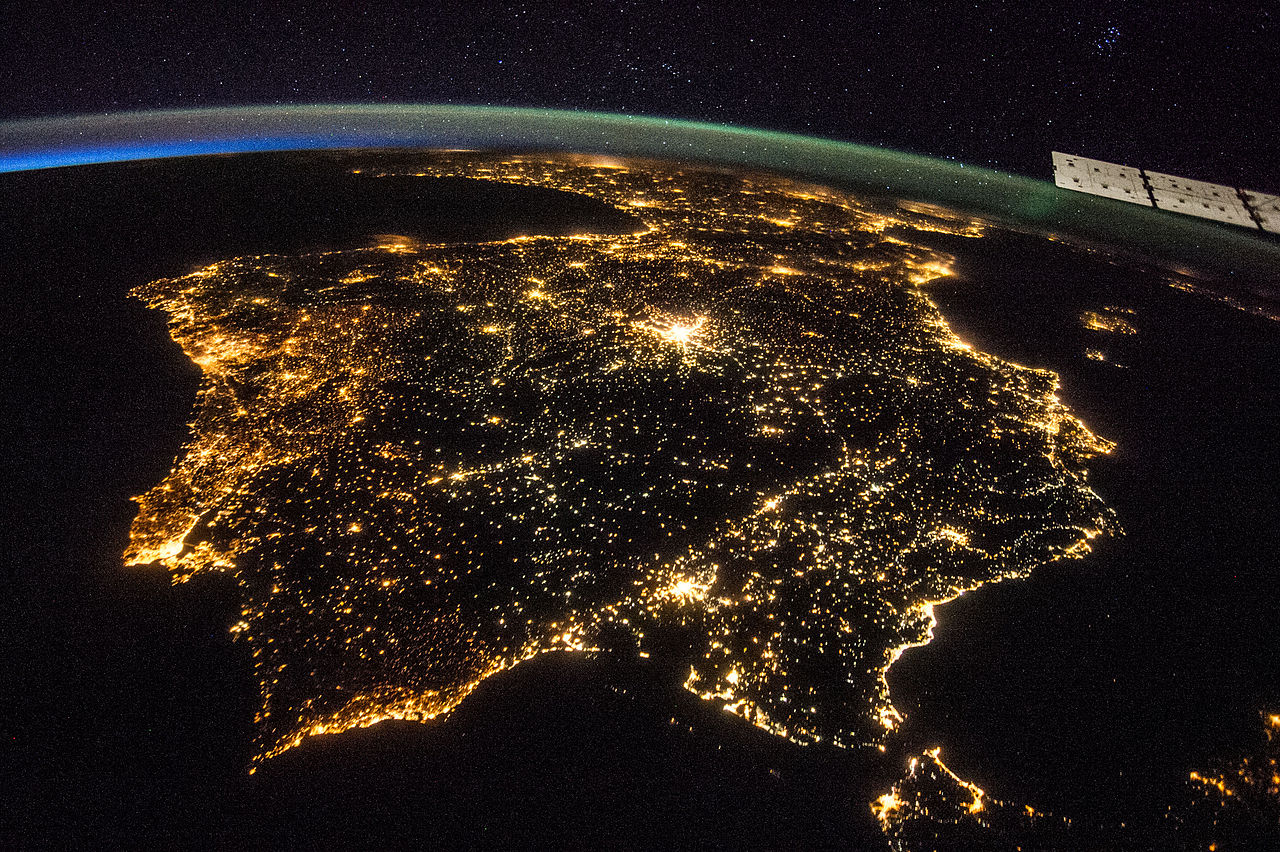}}
 \caption{The Iberian Peninsula at night seen from the International Space 
Station (Image courtesy of the Earth Science and Remote Sensing Unit, NASA 
Johnson Space Center, mission-roll-frame no. ISS040-E-081320 (26~July~2014), 
\url{http://eol.jsc.nasa.gov/SearchPhotos/photo.pl?mission=ISS040&roll=E&frame=081320}).}
 \label{f:iss}
\end{figure}

Navigation using celestial objects is a skill that emerged already long before 
humans roamed the Earth. Today, we know numerous examples among animals who 
find their course using the day or night sky. Bees and monarch butterflies 
navigate by the Sun \citep{sauman_connecting_2005}, just like starlings do 
\citep{kramer_experiments_1952}. Even more impressive is the ability of birds 
\citep{emlen_celestial_1970,lockley_animal_1967,sauer_celestial_1958} and seals 
\citep{mauck_harbour_2008} to identify the position of stars during night-time 
for steering a course. However, in our modern civilisation with intense 
illumination of cities, strong lights can be mistaken for celestial objects. For 
instance, moths use the moon to maintain a constant course, but if confused by a 
street lamp, they keep on circling around it until exhaustion 
\citep{stevenson_probing_2008}. Hence, light pollution is a serious threat to 
many animals. Its magnitude is demonstrated by Fig.~\ref{f:iss}.

Among the first humans to have navigated the open sea were the aboriginal 
settlers of Australia some 50,000 years ago \citep{hiscock_occupying_2013}. The 
oldest records of seafaring in the Mediterranean date back to 7,000 BCE 
\citep{hertel_geheimnis_1990}, admittedly done with boats or small ships that 
were propelled by paddles only. The routes were restricted close the coast where 
landmarks helped to navigate to the desired destinations. In order to be able to 
cross larger distances, propulsion independent of muscle force is needed. 
Therefore, the sail was one of the most important inventions in human history, 
in its significance similar to the wheel. Around the middle of the 4th 
millennium BCE, Egyptian ships sailed the eastern Mediterranean 
\citep{bohn_geschichte_2011} and established trade routes with Byblos in 
Phoenicia, the biblical Canaan, now Lebanon. This is about the time when the 
Bronze Age began. Tin is an important ingredient to bronze. After the depletion 
of the local deposits, tin sources in central and western Europe triggered large 
scale trade \citep{penhallurick_tin_1986}. Transportation over large distances 
inside and outside the Mediterranean was accomplished by ships.

\begin{figure}[!ht]
 \resizebox{\hsize}{!}{\includegraphics{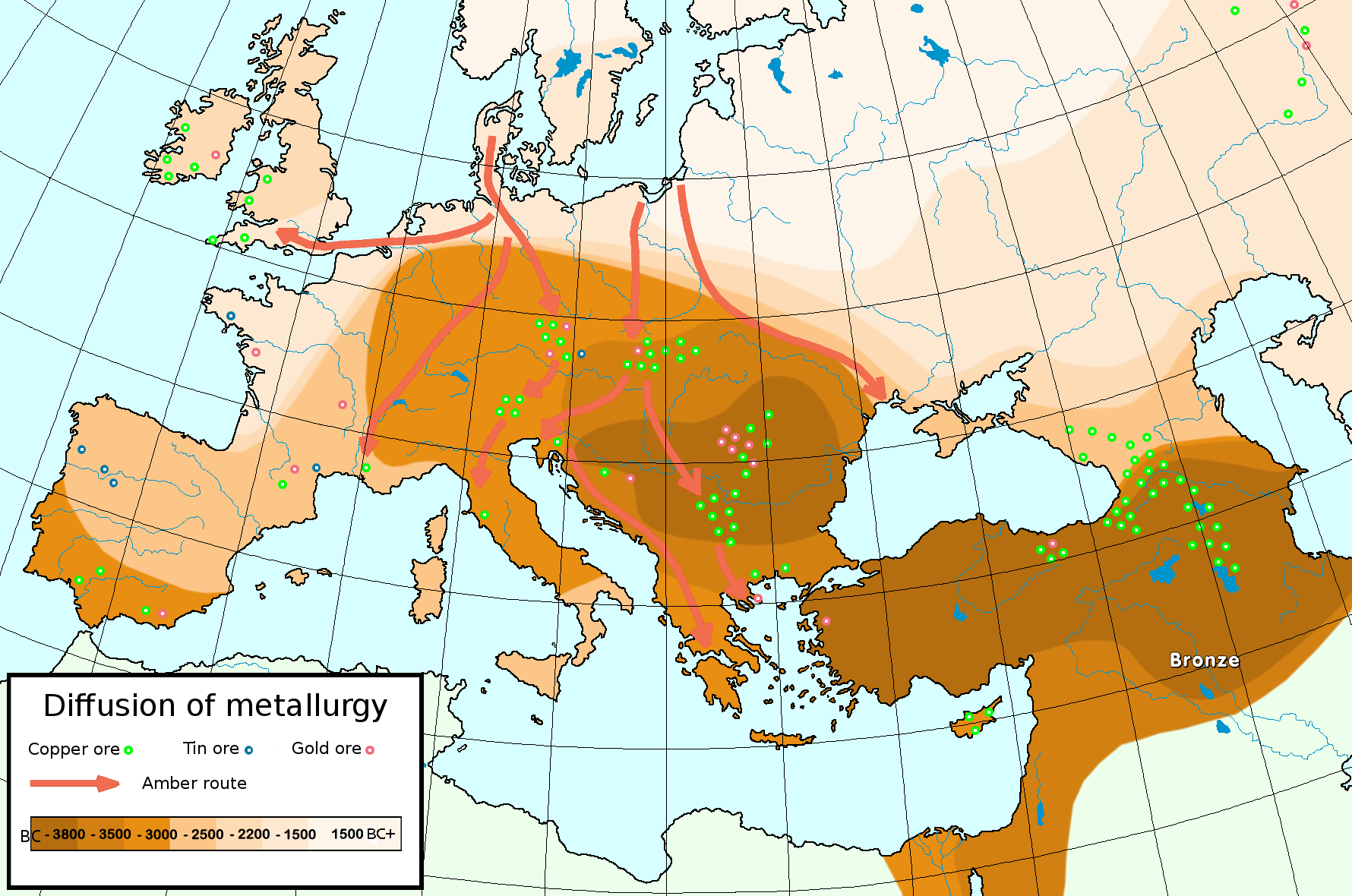}}
 \caption{Map of the diffusion of metallurgy. Bronze Age tin deposits are mostly 
found at the European Atlantic coast (User:Hamelin de Guettelet, 
\url{https://commons.wikimedia.org/wiki/File:Metallurgical_diffusion.png}, 
public domain).}
 \label{f:metallurgy}
\end{figure}

Soon, the navigators realised that celestial objects, especially stars, can be 
used to keep the course of a ship. Such skills have been mentioned in early 
literature like Homer’s Odyssey which is believed to date back to the 8th 
century BCE. The original sources are thought to originate from the Bronze Age, 
in which the Minoans of Crete were a particularly influential people that lived 
between 3,650 and 1,450 BCE in the northern Mediterranean, and who sailed the 
Aegean Sea. Since many of their sacral buildings were aligned with the cardinal 
directions and astronomical phenomena like the rising Sun and the equinoxes 
\citep{henriksson_orientations_2008,henriksson_solar_2009}, it is reasonable to 
think that they used this knowledge for navigation, too 
\citep{blomberg_evidence_1999}. The Minoans sailed to the island of Thera and 
Egypt, which would have taken them on open water for several days.

\begin{figure}
 \resizebox{\hsize}{!}{\includegraphics{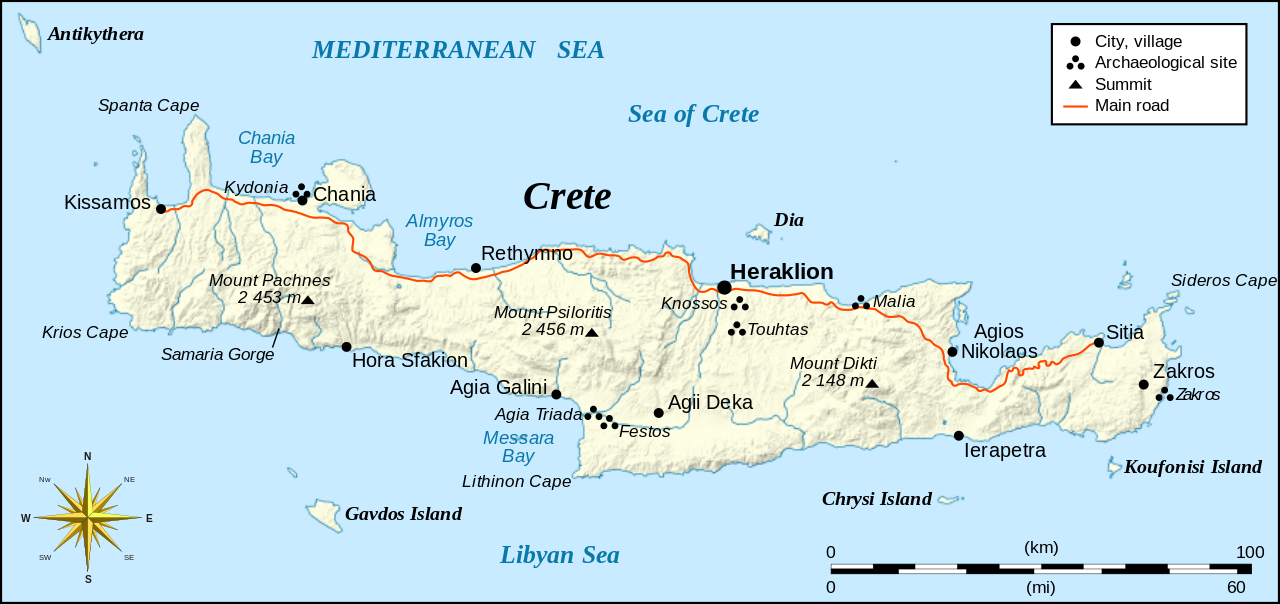}}
 \caption{Map of Crete with ancient Minoan sites in the early 2nd millennium BCE 
(Eric Gaba (Sting), 
\url{https://commons.wikimedia.org/wiki/File:Crete_integrated_map-en.svg}, 
\url{https://creativecommons.org/licenses/by-sa/4.0/legalcode}).}
 \label{f:crete}
\end{figure}

The Greek poet Aratos of Soli published his Phainomena around 275 BCE 
\citep{aratus_callimachus:_1921} in which he provided detailed positions of 
constellations and their order of rising and setting, which would be vital 
information for any navigator to maintain a given course. He would simply have 
pointed his ship at a bearing and be able to keep it with the help of stellar 
constellations that appeared towards that heading. The azimuth of a given star 
when rising or setting remains constant throughout the year, except for a slow 
variation caused by the 26,000 years period of the precession of the Earth's 
axis. Interestingly, Aratos’ positions did not fit the Late Bronze and Early 
Iron Age but the era of the Minoan reign \citep{blomberg_evidence_1999} some 
2,000 years earlier.

\begin{figure}[!ht]
 \resizebox{\hsize}{!}{\includegraphics{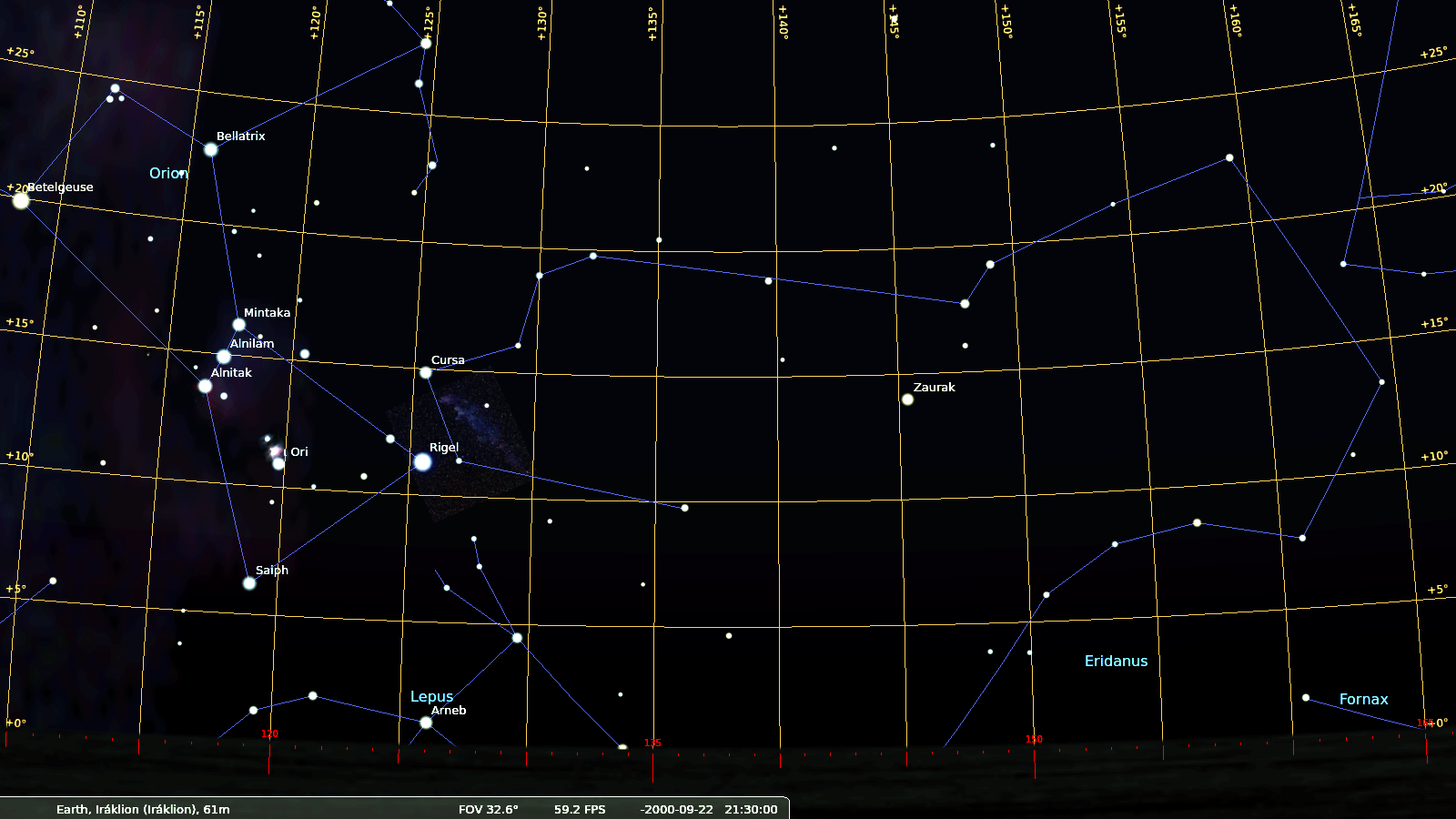}}
 \caption{The night sky with bearing from Crete to Alexandria for 22 September 
2000~BCE, 21:30 UT (own work, created with Stellarium, free GNU GPL software, 
after \citet{blomberg_evidence_1999}, Fig.~9).}
 \label{f:crete2alex}
\end{figure}

Around 1200 BCE, the Phoenicians became the dominating civilisation in the 
Mediterranean. They built colonies along the southern and western coasts of the 
Mediterranean and beyond. Among them was the colony of Gades (now Cadíz) just 
outside the Strait of Gibraltar which served as a trading point for goods and 
resources from Northern Europe 
\citep{cunliffe_extraordinary_2003,hertel_geheimnis_1990}. Several documented 
voyages through the Atlantic Ocean took them to Britain and even several hundred 
miles south along the African coast \citep{johnson_history_2009}.

\begin{figure}[!ht]
 \resizebox{\hsize}{!}{\includegraphics{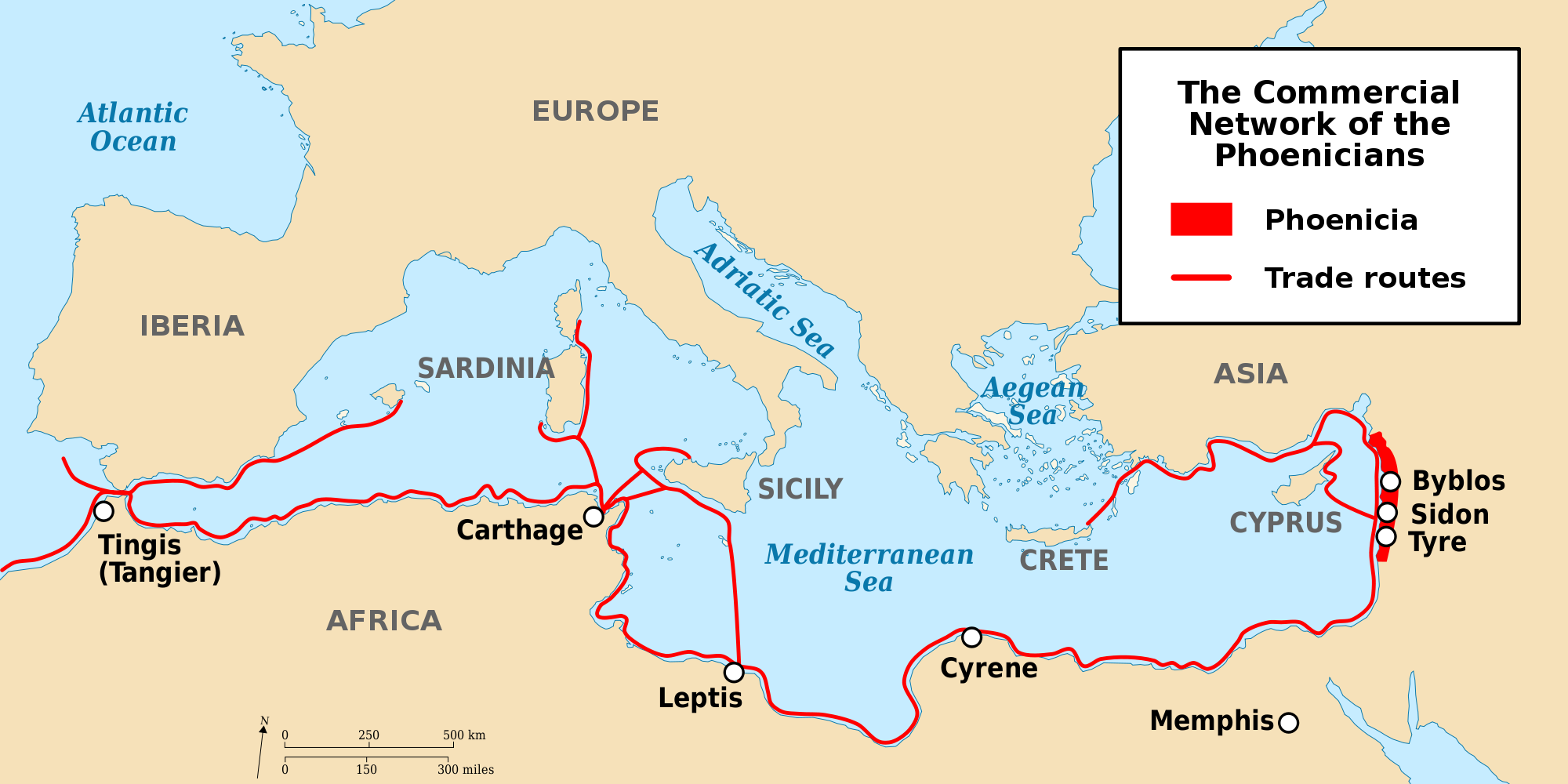}}
 \caption{Trade routes of the Phoenicians during the European Bronze Age (DooFi, 
\url{https://commons.wikimedia.org/wiki/File:PhoenicianTrade_EN.svg}, 
\url{https://creativecommons.org/licenses/by-sa/3.0/legalcode}).}
 \label{f:trade}
\end{figure}

The Greek historian Herodotus (ca.~484 -- 420 BCE) reports of a Phoenician 
expedition funded by the Egyptian Pharaoh Necho II (610 -– 595~BCE) that set out 
from the Red Sea to circumnavigate Africa and returned to Egypt via the 
Mediterranean 
\citep{bohn_geschichte_2011,hertel_geheimnis_1990,johnson_history_2009}. The 
sailors apparently reported that at times the Sun was located North 
\citep{cunliffe_extraordinary_2003} which is expected after crossing the equator 
to the south. All this speaks in favour of extraordinary navigational skills. 
After the Persians conquered the Phoenician homeland in 539~BCE, their influence 
declined, but was re-established by descendants of their colonies, the 
Carthaginians.

\subsection{Pytheas}

\begin{figure}[!ht]
 \centering
 \resizebox{0.4\hsize}{!}{\includegraphics{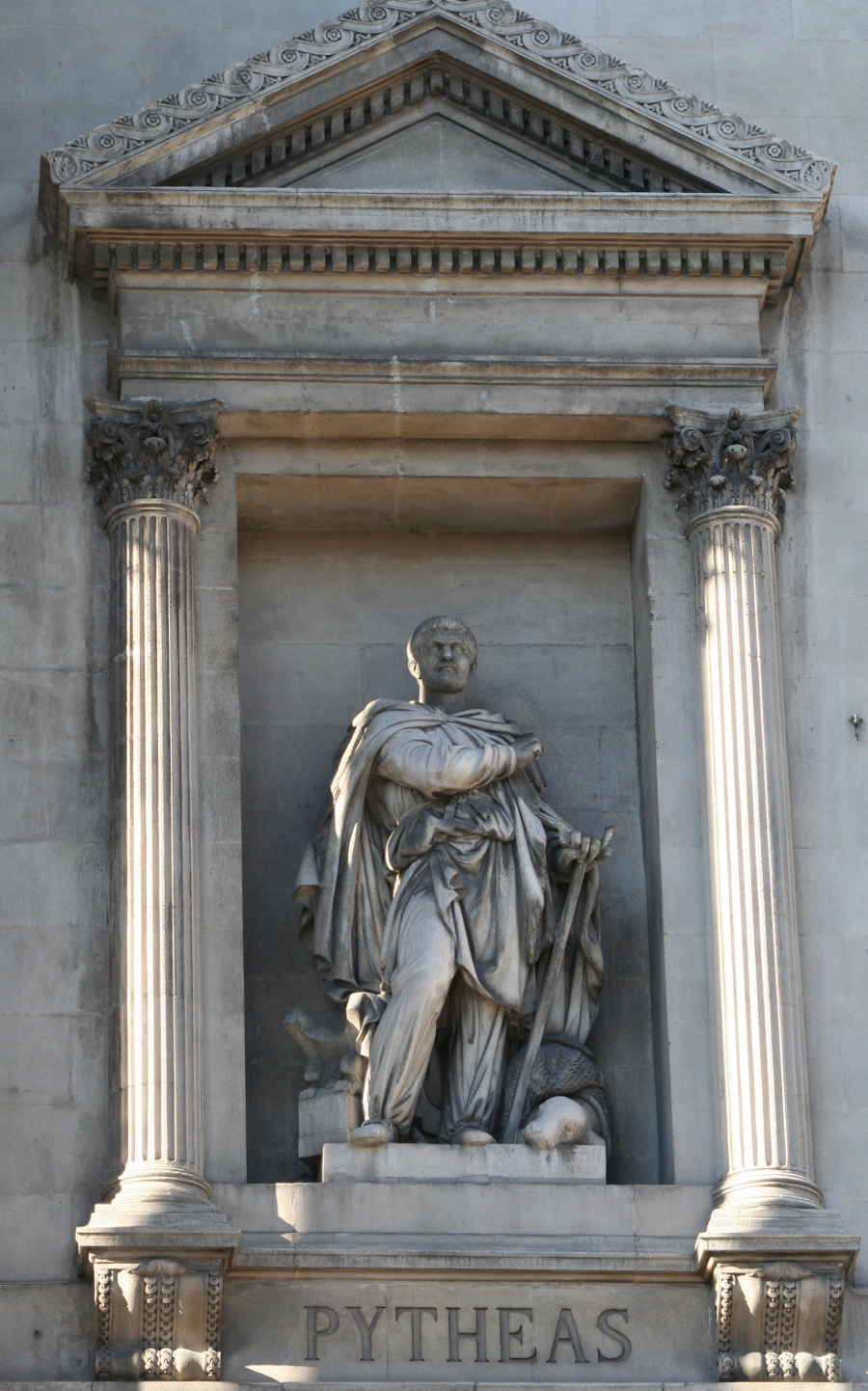}}
 \caption{Statue of Pytheas, erected at the Palais de la Bourse in Marseille in 
honour of his achievements (Rvalette, 
\url{https://commons.wikimedia.org/wiki/File:Pythéas.jpg}, ``Pythéas'', 
\url{https://creativecommons.org/licenses/by-sa/3.0/legalcode}).}
 \label{f:pytheas}
\end{figure}

A very notable and well documented long distance voyage has been passed on by 
ancient authors and scholars like Strabo, Pliny and Diodorus of Sicily. It is 
the voyage of Pytheas (ca.~380 -- 310~BCE), a Greek astronomer, geographer and 
explorer from Marseille who around 320~BCE apparently left the Mediterranean, 
travelled along the European west coast and made it up north until the British 
Isles and beyond the Arctic Circle, during which he possibly reached Iceland or 
the Faroe Islands that he called Thule 
\citep{baker_ancient_1997,cunliffe_extraordinary_2003,hergt_nordlandfahrt_1893}.

Massalia (or Massilia), as it was called then, was founded by Phocean Greeks 
around 600~BCE, and quickly evolved into one of the biggest and wealthiest Greek 
outposts in the Western Mediterranean with strong trade relations to Celtic 
tribes who occupied most of Europe \citep{cunliffe_extraordinary_2003}. Pytheas 
was born into the Late Bronze Age, when the trade with resources from Northern 
Europe was flourishing. Not much was known in Greek geography about this part of 
the world, except that the barbarians living there mined the tin ore and 
delivered the precious amber that the whole Mediterranean so desperately longed 
for. Perhaps it was out of pure curiosity why Pytheas set out to explore these 
shores.

\begin{figure}[!ht]
 \centering
 \resizebox{\hsize}{!}{\includegraphics{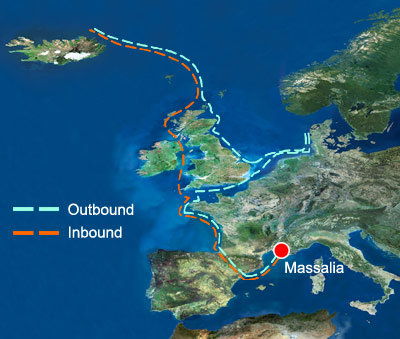}}
 \caption{The journey of Pytheas of Massalia according to 
\citet{cunliffe_extraordinary_2003} (ESA/Cunliffe, 
\url{http://www.esa.int/spaceinimages/Images/2005/09/The_journey_of_Pytheas}, 
\url{http://www.esa.int/spaceinimages/ESA_Multimedia/Copyright_Notice_Images}).}
 \label{f:voyage}
\end{figure}

His voyage was a milestone, because Pytheas was a scientist and a great 
observer. He already used a gnomon or a sundial, which allowed him to determine 
his latitude and measure the time during his voyage 
\citep{nansen_northern_1911}. He also noticed that in summer the Sun shines 
longer at higher latitudes. In addition, he was the first to notice a relation 
between the tides, which are practically not present in the Mediterranean, and 
the lunar phases \citep{roller_through_2006}.

\newglossaryentry{apparent}
{name={Apparent movement},
description={Movement of celestial objects which in fact is caused by the rotation of the Earth.}
}

\newglossaryentry{cardinal}
{name={Cardinal directions},
description={Main directions, i.e. North, South, West, East}
}

\newglossaryentry{circumpol}
{name=Circumpolar,
description={Property of celestial objects that never set below the horizon.}
}

\newglossaryentry{culminate}
{name=Culmination,
description={Passing the meridian of celestial objects. These objects attain their highest or lowest elevation there.}
}

\newglossaryentry{diurnal}
{name=Diurnal,
description={Concerning a period that is caused by the daily rotation of the Earth around its axis.}
}

\newglossaryentry{ele}
{name=Elevation,
description={Angular distance between a celestial object and the horizon.}
}

\newglossaryentry{gc}
{name={Great circle},
description={A circle on a sphere, whose radius is identical to the radius of the sphere.}
}

\newglossaryentry{meridian}
{name=Meridian,
description={A line that connects North and South at the horizon via the zenith.}
}

\newglossaryentry{poleheight}
{name={Pole height},
description={Elevation of a celestial pole. Its value is identical to the latitude of the observer on Earth.}
}

\newglossaryentry{precess}
{name=Precession,
description={Besides the rotation of a gyroscope or any spinning body, the rotation axis often also moves in space. This is called precession. As a result, the rotation axis constantly changes its orientation and points to different points in space. 
The full cycle of the precession of the Earth's axis takes roughly 26,000 years.}
}

\newglossaryentry{spc}
{name={Spherical polar coordinates},
description={The natural coordinate system of a flat plane is Cartesian and measures distances in two perpendicular directions (ahead, back, left, right). For a sphere, this is not very useful, because it has neither beginning nor ending. Instead, the fixed point is the centre of the sphere. When projected outside from the central position, any point on the surface of the sphere can be determined by two angles with one of them being related to the symmetry axis. Such axis defines two poles. In addition, there is the radius that represents the third dimension of space, which permits determining each point within a sphere. This defines the spherical polar coordinates. When defining points on the surface of a sphere, the radius stays constant.}
}

\newglossaryentry{sundial}
{name=Sundial,
description={A stick that projects a shadow cast by the Sun. The orientation and length of the shadow permits determining time and latitude.}
}

\newglossaryentry{zenith}
{name=Zenith,
description={Point in the sky directly above.}
}

\section{List of material}
The list contains items needed by one student. The teacher may decide that they 
work in groups of two.

\begin{itemize}
\item Worksheet
\item Pair of compasses
\item Pencil
\item Ruler
\item Calculator
\item Protractor
\item Torch (for supplemental third activity)
\item Magnetic compass (optional, third activity)
\item Computer with MS Excel installed
\item Excel spreadsheet:\\
astroedu1645\_AncientMediterranean\_BrightStars.xlsx
\end{itemize}

\section{Goals}
With this activity, the students will learn that
\begin{itemize}
\item celestial navigation has been developed already many centuries ago.
\item apart from using Polaris there are other methods to determine cardinal 
directions from the positions of stars.
\item ancient navigators were able to successfully navigate on open water 
following stars and constellations.
\end{itemize}

\section{Learning objectives}
The students  will be able to
\begin{itemize}
\item describe methods to determine the cardinal directions form observing the sky.
\item name prominent stellar constellations.
\item explain the nature of circumpolar stars and constellations.
\item use an Excel spreadsheet for calculations.
\item describe the importance of improved navigational skills for early civilisations.
\end{itemize}

\section{Target group details}

\noindent
Suggested age range: 14 -- 19 years\\
Suggested school level: Middle School, Secondary School\\
Duration: 90 minutes

\section{Evaluation}
According to the suggested questions listed in the description of the activity, 
the teacher should guide the students to recognise the positions and the 
apparent movement of celestial objects as indicators for cardinal directions.

Before working on activity 1, the students should look closely at the map 
provided. A visit at a planetarium helps remembering the constellations. Let the 
students name constellations they already know.

Ask the students (see Q\&A in the activity description) where the North Star 
would be when observed from the terrestrial North Pole and the Equator. Then ask 
them, how this position changes when travelling between these locations. When 
that concept is understood, introduce the rotation and the apparent motion of 
the stars. Show them the picture of the star trails and ask them, where they 
come from. Ask them, which of the stars or constellations remain above the 
horizon for the different locations on Earth mentioned above. Those are 
circumpolar stars and constellations.

Explain the usage of the Excel spreadsheet needed for activity 2. Let the 
students compare their results for different latitudes.

Discuss with the students, what the reasons for seafaring could have been in 
ancient epochs.

The third optional activity acts as a wrap-up and can be used to evaluate, what 
the students have understood.

\section{Full description of the activity}
\subsection{Introduction}
It would be beneficial, if the activity be included into a larger context of 
seafaring, e.g. in geography, history, literature, etc.

Tip: This activity could be combined with other forms of acquiring knowledge 
like giving oral presentations in history, literature or geography highlighting 
navigation. This would prepare the field in a much more interactive way than 
what a teacher can achieve by summarising the facts.

There are certainly good documentaries available on sea exploration that could 
be shown. As an introduction to celestial navigation in general and the early 
navigators, let the students watch the following videos. The last one is in 
French. This could be done in junction with French lessons in school. If not, 
tell the story about Pytheas as outlined in the background information. A link 
to literature or history classes may be established by reading ``The 
Extraordinary Voyage of Pytheas'' by B. Cunliffe.

\medskip
\noindent
Episode 2: Celestial Navigation (Duration: 4:39)\\
\url{https://www.youtube.com/watch?v=DoOuSo9qElI}

\medskip
\noindent
How did early Sailors navigate the Oceans? | The Curious Engineer (Duration: 6:20)\\
\url{https://www.youtube.com/watch?v=4DlNhbkPiYY}

\medskip
\noindent
World Explorers in 10 Minutes (Duration: 9:59)\\
\url{https://www.youtube.com/watch?v=iUkOfzhvMMs}

\medskip
\noindent
Once upon a time … man: The Explorers - The first navigators (Duration: 23:13)\\
\url{https://www.youtube.com/watch?v=KuryXLnHsEY}

\medskip
\noindent
Pythéas, un Massaliote méconnu (French, duration: 9:57)\\
\url{https://www.youtube.com/watch?v=knBNHbbu-ao}

\medskip
Ask the students, if they had an idea for how long mankind already uses ships to 
cross oceans. One may point out the spread of the Homo sapiens to islands and 
isolated continents like Australia.

\medskip
\noindent
\textit{Possible answers:}\\
We know for sure that ships have been used to cross large distances already 
since 3,000 BCE or earlier. However, the early settlers of Australia must have 
found a way to cross the Oceans around 50,000~BCE.

\medskip
Ask them, what could have been the benefit to try to explore the seas. Perhaps, 
someone knows historic cultures or peoples that were famous sailors. The teacher 
can support this with a few examples of ancient seafaring peoples, e.g. from the 
Mediterranean.

\medskip
\noindent
\textit{Possible answers:}\\
Finding new resources and food, trade, spirit of exploration, curiosity.

\begin{table*}
\caption{List of cities along with their latitudes. The solutions from activity 
1 are added in italic writing.}
\label{t:a1}
\centering
\begin{tabular}{c c c l}
\hline\hline
City & Latitude ($\degr$) & Radius in map (cm) & \multicolumn{1}{c}{Constellations}\\
\hline
Tunis (ancient & \raisebox{-6pt}{36.8} & \raisebox{-6pt}{3.7} & 
\raisebox{-6pt}{\em Ursa Minor, Ursa Major (Big Dipper), Draco, Cepheus, 
Cassiopeia}\\[-4pt]
Carthage, Tunisia) & & & \\[4pt]
Cape Town & \raisebox{-6pt}{-33.9} & \raisebox{-6pt}{3.4} & \em Crux, Pavo 
(Peacock), Achernar, most of Carina, Toliman \\[-4pt]
(South Africa) & & & \em of Centaurus\\[4pt]
Plymouth & \raisebox{-6pt}{50.4} & \raisebox{-6pt}{5.0} & \em Ursa Minor, Ursa 
Major, Draco, Cepheus, Cassiopeia, Perseus\\[-4pt]
(UK) & & & \em (mostly), part of Cygnus with Deneb\\[4pt]
Wellington & \raisebox{-6pt}{-41.3} & \raisebox{-6pt}{4.1} & \raisebox{-6pt}{\em 
Carina (incl. Canopus), Ara, Pictor, Dorado}\\[-4pt]
(New Zealand) & & & \\[8pt]
Mumbai & \raisebox{-6pt}{19.0} & \raisebox{-6pt}{1.9} & \raisebox{-6pt}{\em Ursa 
Minor}\\[-4pt]
(India) & & & \\[4pt]
Grytviken & \raisebox{-6pt}{-54.3} & \raisebox{-6pt}{5.4} & \raisebox{-6pt}{\em 
Centaurus, Lupus, Main part of Puppis, Phoenix, Grus}\\[-4pt]
(South Georgia) & & & \\
\hline
\end{tabular}
\end{table*}

\medskip
Ask the students, how they find the way to school every day. What supports their 
orientation to not get lost? As soon as reference points (buildings, traffic 
lights, bus stops, etc.) have been mentioned ask the students how navigators 
were able to find their way on the seas. In early times, they used sailing 
directions in connection to landmarks that can be recognised. But for this, the 
ships would have to stay close to the coast. Lighthouses improved the situation. 
Magnetic compasses have been a rather late invention around the 11th century CE, 
and they were not used in Europe before the 13th century. But what could be used 
as reference points at open sea? Probably the students will soon mention 
celestial objects like the Sun, the Moon and stars.

Suggested additional questions, especially after showing the introductory videos:

\medskip
\noindent
Q: Who was Pytheas?\\\noindent
A: He was an ancient Greek scientist and explorer.

\medskip
\noindent
Q: Where and when did he live?\\\noindent
A: He lived in the 4th century BCE during the Late Bronze Age in Massalia, now 
Marseille.

\medskip
\noindent
Q: Where did he travel?\\\noindent
A: Pytheas travelled north along the Atlantic coast of Europe to Britain and 
probably to the Arctic Circle and Iceland.

\medskip
\noindent
Q: What did he observe and discover during his voyage?\\\noindent
A: He was the first Greek to travel so far to the north. He noticed that the 
length of daylight depends on latitude. He was also the first to relate the 
tides to the phases of the moon.

\subsection{Activity 1: Circumpolar constellations and stars}
In absence of a bright star at the celestial poles, ancient navigators were able 
to find it by observing a few circumpolar stars. These navigators were 
experienced enough to determine true north by recognising the relative position 
of such stars and by their paths around it.

In addition, they used circumpolar constellations and stars to infer their 
latitude. This means, they never rise or set – they are always above the 
horizon. While today we can simply measure the elevation of Polaris above the 
horizon, ancient navigators saw that star many degrees away from the celestial 
North Pole. In the southern hemisphere, there is no such stellar indicator 
anyway. So, instead of measuring the elevation of Polaris, they observed which 
stars and constellations were still visible above the horizon when they attained 
their lowest elevation above the horizon (lower culmination) during their 
apparent orbit around the celestial pole.

Let the students watch the two following videos that demonstrate the phenomenon 
of circumpolar stars and constellations for two locations on Earth. They show 
the simulated daily apparent rotation of the sky around the northern celestial 
pole.

\medskip
\noindent
CircumpolarStars Heidelberg 49degN (Duration: 0:57)\\\noindent
\url{https://youtu.be/uzeey9VPA48}

\medskip
\noindent
CircumpolarStars Habana 23degN (Duration: 0:49)\\\noindent
\url{https://youtu.be/zggfQC_d7UQ}

\medskip\noindent
The students will notice that
\begin{enumerate}
\item there are always stars and constellations that never set. Those are the 
circumpolar stars and constellations.
\item the angle between the celestial pole (Polaris) and the horizon depends on 
the latitude of the observer. In fact, these angles are identical.
\item the circumpolar region depends on the latitude of the observer. It is 
bigger for locations closer to the pole.
\end{enumerate}

If the students are familiar with the usage of a planisphere, they can study the 
same phenomenon by watching the following two videos.

\medskip
\noindent
CircumPolarStars phi N20 (Duration: 0:37)\\\noindent
\url{https://youtu.be/Uv-xcdqhV00}

\medskip
\noindent
CircumPolarStars phi N45 (Duration: 0:37)\\\noindent
\url{https://youtu.be/VZ6RmdzbpPw}

\medskip
They show the rotation of the sky for the latitudes $20\degr$ and $45\degr$. The 
transparent area reveals the visible sky for a given point in time. The dashed 
circle indicates the region of circumpolar stars and constellations.

\begin{figure*}
 \centering
 \resizebox{\hsize}{!}{\includegraphics{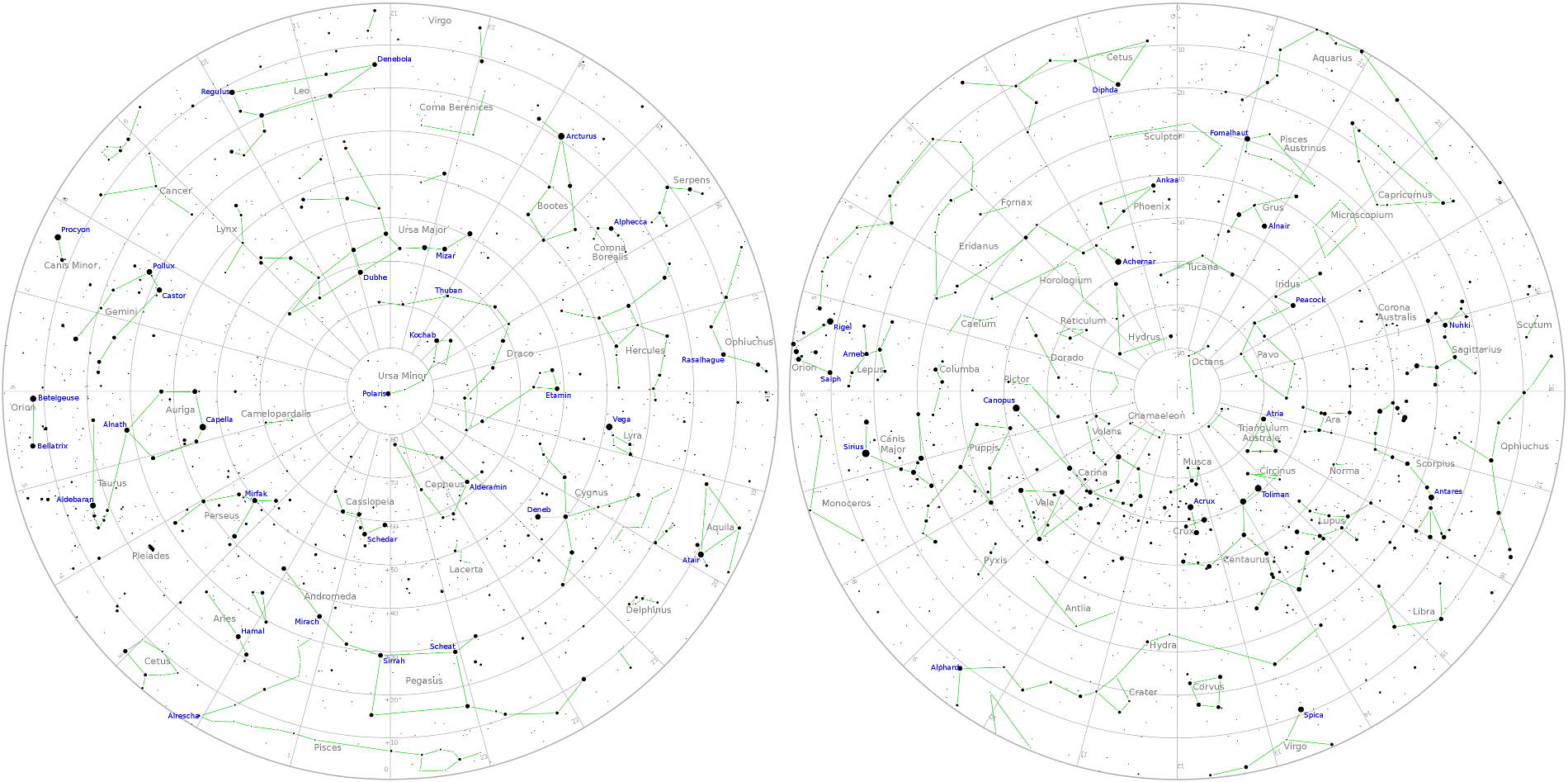}}
 \caption{Star charts of the northern and the southern hemisphere (Markus 
Nielbock, 
\url{https://commons.wikimedia.org/wiki/File:NorthernCelestialHemisphere.png}, 
\url{https://commons.wikimedia.org/wiki/File:SouthernCelestialHemisphere.png}, 
\url{https://creativecommons.org/licenses/by/4.0/legalcode}, created with PP3, 
\url{http://pp3.sourceforge.net})}
 \label{f:hemispheres}
\end{figure*}

\subsubsection{Questions}
\noindent
Q: What is special about the geographic North and South Poles of the Earth 
compared to other locations?\\\noindent
A: They define the rotation axis of the Earth.

\medskip
\noindent
Q: How do you find North and the other cardinal directions without a compass?
A: Celestial bodies, e.g. stars like Polaris which indicates the celestial north pole.

\medskip
\noindent
Q: Why does the North Star (Polaris) indicate North?\\\noindent
A: In our lifetime, Polaris is close to the celestial north pole.

\medskip
\noindent
Q: Where in the sky would be the celestial North/South Pole if you stood exactly 
at the terrestrial North/South Pole?\\\noindent
A: At the zenith, i.e. directly overhead.

\medskip
\noindent
Q: How would this position change, if you travelled towards the equator?\\\noindent
A: Its elevation would decline from zenith to the horizon.

\medskip
\noindent
Q: What are circumpolar constellations?\\\noindent
A: These are constellations that revolve around one of the celestial poles and 
never rise or set. They are always above the horizon.

\medskip
\noindent
Q: Which of the visible constellations would be circumpolar, if you stood on the 
terrestrial North/South Pole/equator?\\\noindent
A: The entire northern/southern hemisphere (poles). None at the equator.

\medskip
\noindent
Q: If the North Star was not visible, how would you be able to determine you latitude anyway?\\\noindent
A: Since the circumpolar stars and constellations depend on the latitude, just 
like the elevation of Polaris, the ones that always stay above the horizon 
indicate, where I am.

\subsubsection{Exercise}
The task is now to walk in the footsteps of a navigator that lived around 5,000 
years ago. Based on those skills, the students will determine the constellations 
that are circumpolar when observed from given positions on Earth.

The table below contains the names of six cities along with their latitudes 
$\varphi$. Negative values indicate southern latitudes. A seventh row is empty, 
where the students can add the details of their home town. From this, they will 
have to calculate the angular radii $\varrho$ from the celestial pole. The 
calculation is simple, because is the same as the pole height and the latitude:

$$\varphi = \varrho$$

Then they select the map that matches the hemisphere. The students use the 
compasses to draw circles of those radii around the corresponding pole. The 
constellations inside that circle are circumpolar. The constellations that are 
just fully or partially visible for a given city are added to the table.

Possible solutions are added in italics. The table prepared for the exercise is 
contained in the worksheet.

\subsubsection{Detailed instructions}
\begin{enumerate}
\item Determine the map scale. The angular scale is $90\degr$ from the poles to 
the outer circle, i.e. the celestial equator.
\item Convert the latitudes in the table into radii in the scale of the maps and 
add them to the table.
\item For each of the cities:
\begin{enumerate}
\item Select the suitable map.
\item Use the compasses to draw a circle with a radius that was determined for 
that city.
\item Find and note the visible circumpolar constellations. If they are too 
many, just select the most prominent ones.
\end{enumerate}
\end{enumerate}

\subsubsection{Discussion}
In ancient times, Polaris did not coincide with the celestial North Pole. 
Explain the importance of circumpolar stars and constellations for ancient 
navigators.

\medskip\noindent
Possible result:\\\noindent
They provided an excellent tool to maintain latitude and helped to not get lost 
at open sea.

\subsubsection{Solutions}
The map scale is: $1\,{\rm cm} \propto 10\degr$

\begin{figure*}
 \centering
 \resizebox{\hsize}{!}{\includegraphics{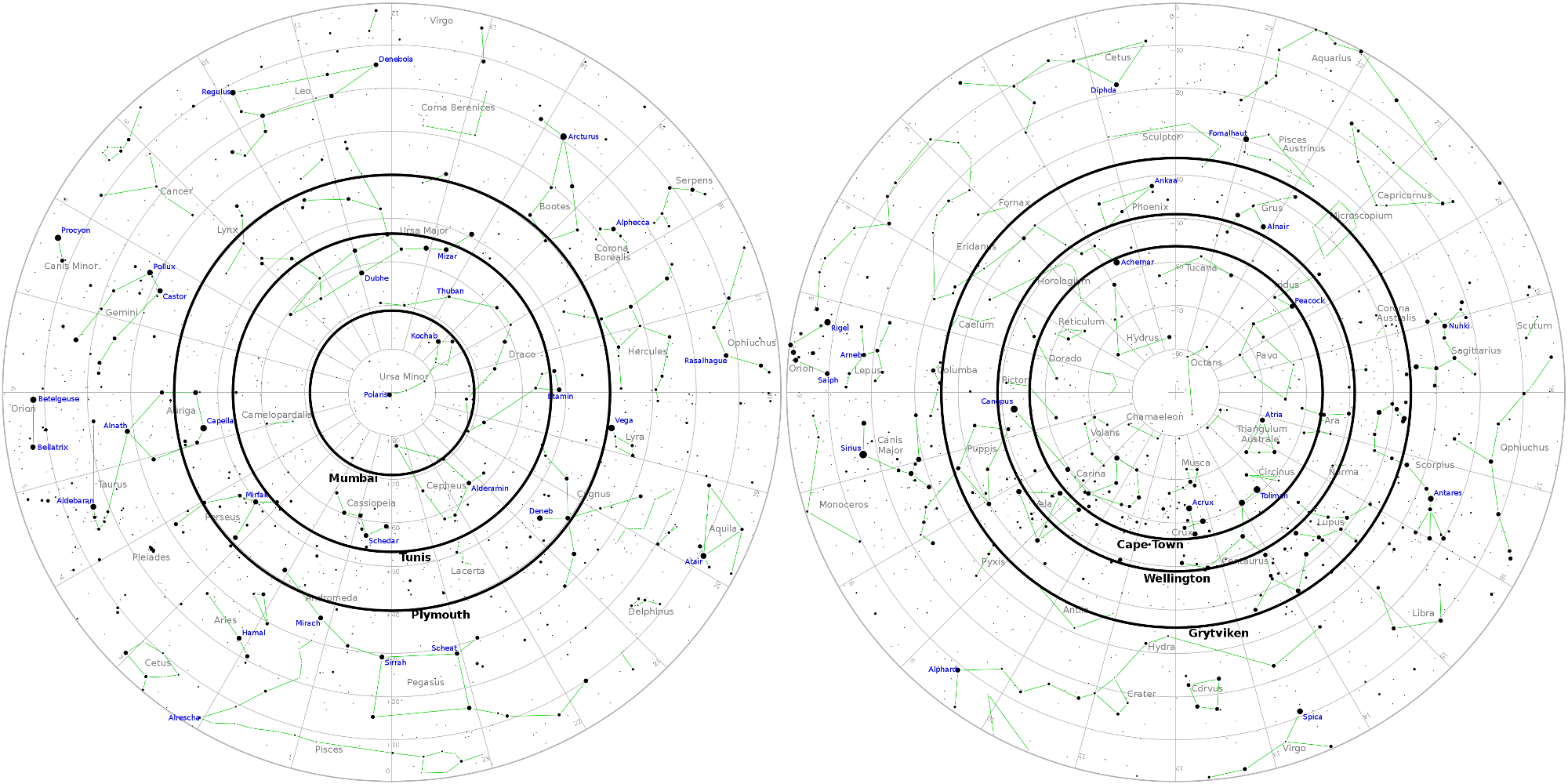}}
 \caption{Solution for the northern and southern locations.}
 \label{f:solutions}
\end{figure*}

\subsection{Activity 2: Stars guide the way}
In the absence of a star like Polaris that indicates a celestial pole, ancient navigators used other stars and constellations to determine cardinal directions and their ship’s course. They realised that the positions where they appear and disappear at the horizon (the bearings) do not change during a lifetime. Experienced navigators knew the brightest stars and constellations by heart.

\subsubsection{Questions}
\noindent
Q: Can you determine the cardinal directions from other stars than Polaris? Note that there is no star at the South Pole.\\\noindent
A: Yes. If you know the stars and constellations, they can guide the way, as they return to the same positions each day.

\medskip\noindent
Q: Why can you use rising and setting stars and constellations to steer a course on sea?\\\noindent
A: The position at the horizon when rising and setting does not change (except for a very slow long term variation).

\medskip\noindent
Q: Would you be able to see the same stars every night during the year?\\\noindent
A: No, the time of rise and set changes. Stars visible during winter nights are up during summer daytimes.

\begin{figure}[!b]
 \centering
 \resizebox{0.4\hsize}{!}{\includegraphics{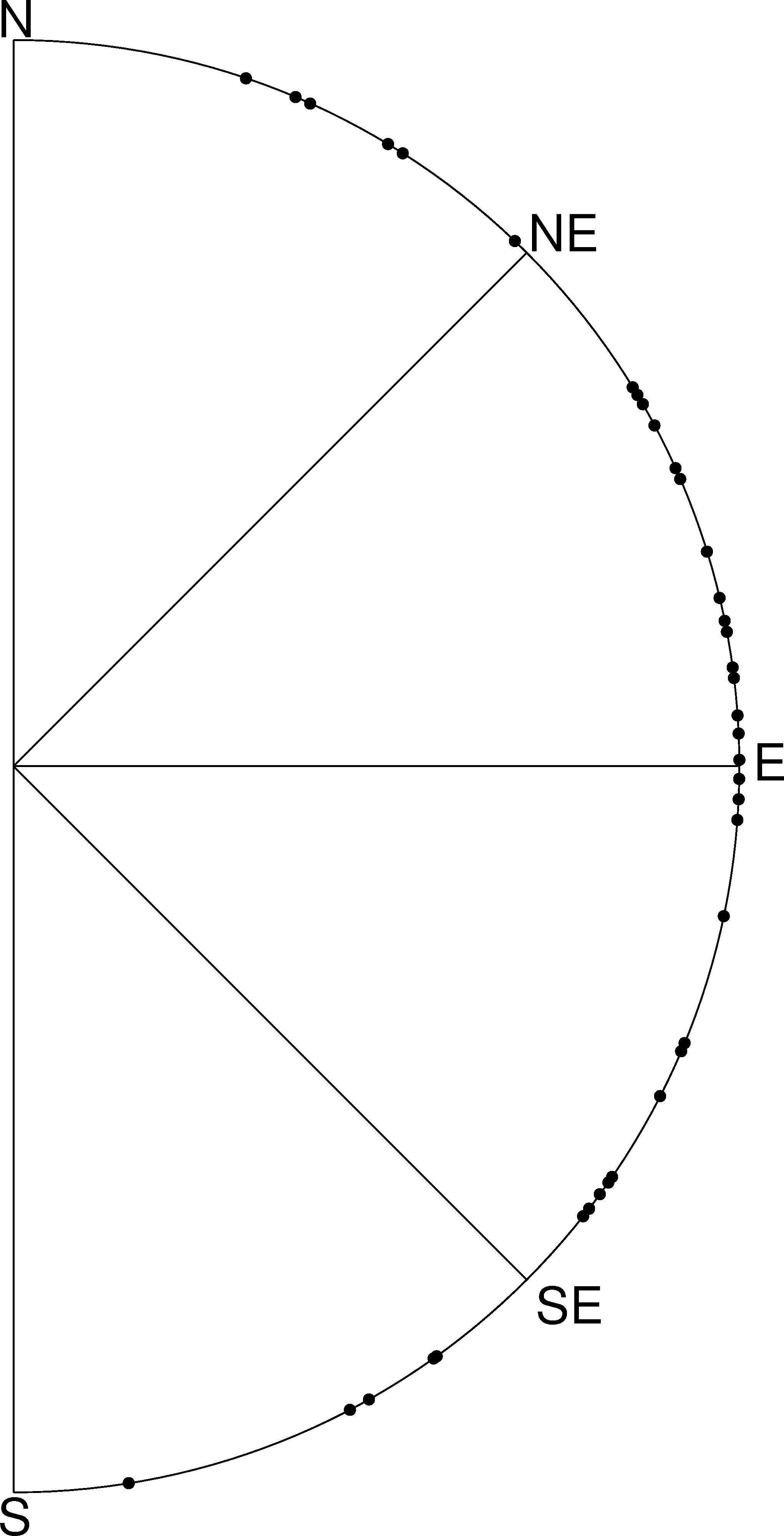}}
 \caption{Bearings of selected rising bright stars for latitude $45\degr$ at an elevation of $10\degr$ above the horizon (own work).}
 \label{f:stellarcompass}
\end{figure}

\subsubsection{Exercise}
The students will produce a stellar compass similar to Fig.~\ref{f:stellarcompass}. The calculations that are needed to convert the sky coordinates of the stars into horizontal coordinates, i.e. azimuth and elevation, are pretty complex. Therefore, this activity comes with an Excel file that does it for them. It consists of 57 bright stars plus the Pleiades which is a very prominent group of stars.

All they have to do is entering the latitude of their location and the elevation of the stars in the corresponding line at the bottom of the spreadsheet. For the elevation, $10\degr$ is a good value. This means, they will get the azimuths of the stars when observed at an elevation of $10\degr$. One can also use different values, but this exercise is meant for finding stars that just rise or set. The azimuth is an angle along the horizon, counting clockwise from North.

\begin{figure}[!b]
 \centering
 \resizebox{0.75\hsize}{!}{\includegraphics{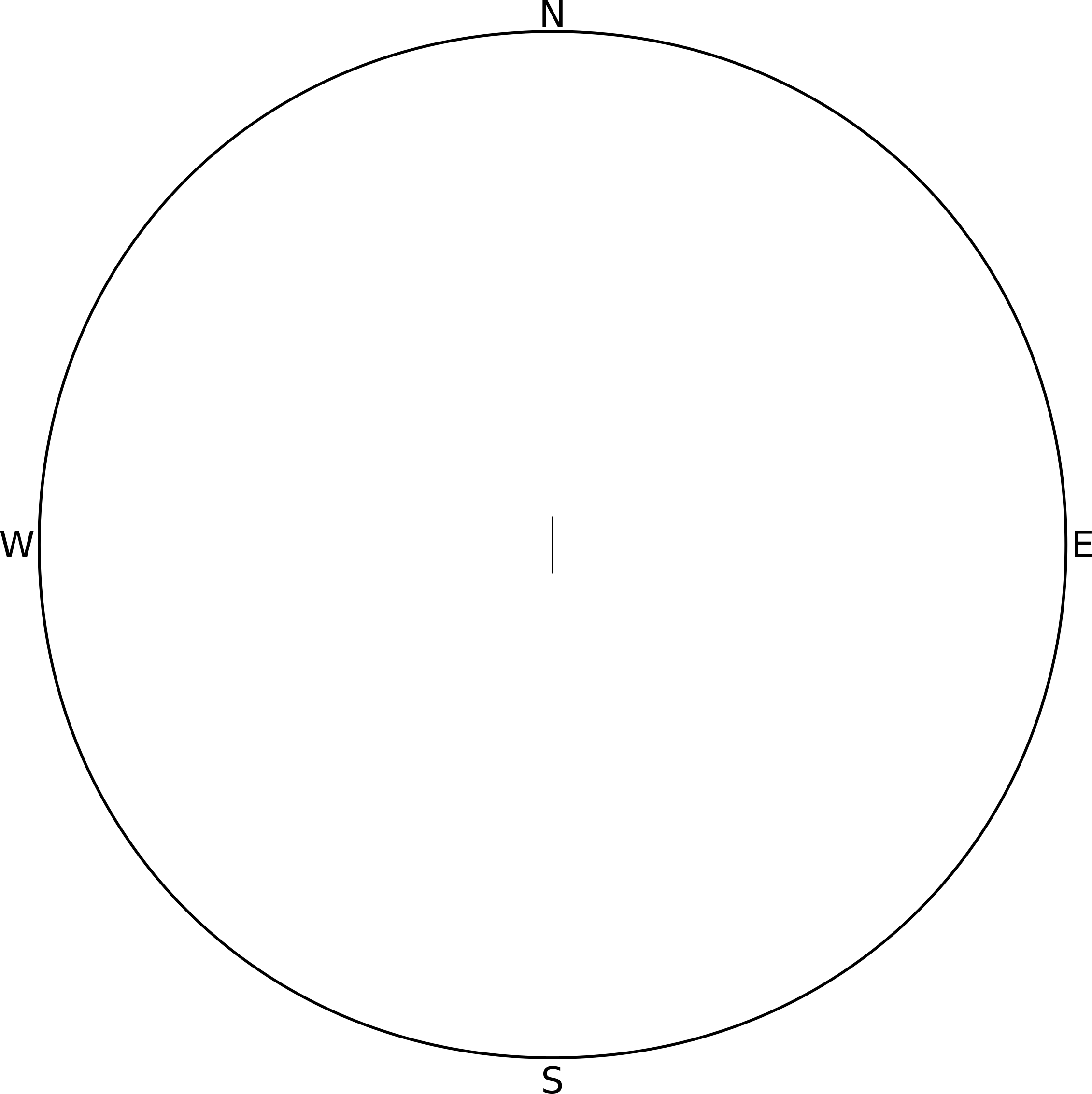}}
 \caption{Students' template for creating their own stellar compass (own work).}
 \label{f:stellarcompass2}
\end{figure}

The last two columns (AZ1, AZ2) then display two azimuths, one when the star is rising and one when the star is setting. Note that the distribution of azimuths for rising and setting stars is symmetric relative to the meridian, i.e. the line that connects North and South. The cells that show \#NA do not contain valid numbers. These stars never rise or set. They are either circumpolar or below the horizon.

The students translate the values into the stellar compass below. They use a protractor and indicate the position of each star on the circle. Then they write its name next to it.

\subsubsection{Discussion}
One of the methods to navigate through the ancient Mediterranean was to stay close to the shores. Besides the danger of shallow waters, explain why Bronze Age mariners must have had methods that would have allowed them to safely navigate on open waters. You may want to look at a map of the Mediterranean.

Possible answers:
The ancient peoples also visited islands for trade or other reasons. Many of them are not visible from coastlines of the Mediterranean. The voyages often would also last longer than just a few hours. Vessels of that age were able to pass five nautical miles per hour on average. There are also reports that were passed on through the ages which tell us about celestial navigation.

\subsection{Activity 3: Do it yourself! (Supplemental)}
Nothing is more instructive than actually applying to real conditions what has been learned and exercised in theory. Therefore, the results from the previous two activities can be tested in the field by observing the night sky.

This activity can be done by the students themselves at home or as a group event with the class.

Select a clear evening and a site with a good view to the horizon. As soon as it is dark enough to see the stars, let the students use their dimmed torches to inspect their maps with the circumpolar ranges from activity 1. A dimmed torch - even better: a red one – helps to keep the eyes adapted to the dark.

After identifying the brightest stars, let them use their stellar compasses from activity 2. The students should point the markers of one or some of the stars to the stars at the sky. Let them identify North (or South, depending on which celestial pole is visible from your location). If in the northern hemisphere, does this match the direction to the North Star, Polaris? In the southern hemisphere, a magnetic compass might be needed.

Let the students identify the constellations they see in the sky on their maps. Ask them to look North (South in the southern hemisphere) and name the stars and constellations that are just above the horizon. Does this coincide with the maps? Note that there should be a circle that indicates the circumpolar range for the local latitude.

Try to highlight that by doing this activity, they are working like the navigators from 4,000 years ago.

\section{Connection to school curriculum}
This activity is part of the Space Awareness category ``Navigation Through The Ages'' and related to the curricula topics:
\begin{itemize}
\item Coordinate systems
\item Basic concepts, latitude, longitude
\item Celestial navigation
\item Constellations
\item Instruments
\end{itemize}

\section{Conclusion}
The students learn about navigational methods and 
seafaring of the ancient epochs like the Bronze Age. Within two activities (plus one optional supplemental activity), they 
will learn how the apparent diurnal paths of stars can help to find the cardinal 
directions and to set course to known destinations in the Mediterranean.

\begin{acknowledgements}
This resource was developed in the framework of Space Awareness. Space Awareness 
is funded by the European Commission’s Horizon 2020 Programme under grant 
agreement no. 638653.
\end{acknowledgements}

\bibliographystyle{aa} 
\bibliography{Navigation}

\glsaddall
\printglossaries

\begin{appendix}
 \section{Supplemental material}
This unit is part of a larger educational package called “Navigation Through the Ages” that introduces several historical and modern techniques used for navigation. An overview is provided via:

\href{http://www.space-awareness.org/media/activities/attach/b3cd8f59-6503-43b3-a9e4-440bf7abf70f/Navigation\%20through\%20the\%20ages\%20compl\_z6wSkvW.pdf}{Navigation\_through\_the\_Ages.pdf}

The supplemental material is available online via the Space Awareness project website at \url{http://www.space-awareness.org}. The direct download links are listed as follows:
 
 \begin{itemize}
  \item Worksheets: \href{https://drive.google.com/file/d/0Bzo1-KZyHftXNDY2bEktbW5kZG8/view?usp=sharing}{astroedu1645-Ancient-Mediterranean-WS.pdf}
  \item Excel file: \href{https://drive.google.com/file/d/0Bzo1-KZyHftXbjYtOE5CbHlkckU/view?usp=sharing}{astroedu1645\_AncientMediterranean\_BrightStars.xlsx}
 \end{itemize}

\end{appendix}

\end{document}